\documentclass[reprint,amssymb, amsmath, aip,cha]{revtex4-1}
\usepackage{graphicx}
\usepackage{amsmath}
\usepackage{bm}%
\usepackage[colorlinks=true,linkcolor=blue]{hyperref}%
\expandafter\ifx\csname package@font\endcsname\relax\else
 \expandafter\expandafter
 \expandafter\usepackage
 \expandafter\expandafter
 \expandafter{\csname package@font\endcsname}%
\fi
\begin{document}

\title{Observation of Radially Inward Turbulent Particle Flux in ETG dominated Plasma of LVPD}%
\author{Prabhakar Srivastav}
\email{prabhakarbhu01@gmail.com}
\affiliation{ Institute For Plasma Research, Gandhinagar-382428, India}%
\affiliation{ Homi Bhabha National Institute, Mumbai-400094, India}%

\author{Rameswar Singh}
\author{L.M. Awasthi}
\email{kushagra.lalit@gmail.com}
\affiliation{ Institute For Plasma Research, Gandhinagar-382428, India}%
\affiliation{ Homi Bhabha National Institute, Mumbai-400094, India}%

\author{A.K. Sanyasi}
\affiliation{ Institute For Plasma Research, Gandhinagar-382428, India}%
\author{P.K. Srivastava}
\affiliation{ Institute For Plasma Research, Gandhinagar-382428, India}%
\author{R. Sughandhi}
\affiliation{ Institute For Plasma Research, Gandhinagar-382428, India}%
\author{R.Singh}
\affiliation{Advance Technology Centre, NFRI, Rep. Korea} 
\author{P.K.Kaw}
\affiliation{ Institute For Plasma Research, Gandhinagar-382428, India}%
\affiliation{ Homi Bhabha National Institute, Mumbai-400094, India}%
\date{\today}
\begin{abstract}
Radially inward turbulent particle flux is observed in the core region of target plasma of Large Volume Plasma Device(LVPD)where electron temperature driven turbulence condition satisfied region satisfy conditions for ETG turbulence, i.e. threshold condition,   $ \eta_e = L_{n_e} / L_{T_e} > 2/3 $ , where density scale length, $ L_{n_e} \sim 300 cm $ and temerature scale length, $ L_{T_e} \sim 50cm $[S.K. Mattoo et al., Phys. Rev. Lett., 108, 255007(2012)\cite{Mattoo_PRL}]. The measured flux is dominantly electrostatic ($\Gamma_{es} \approx 10^{5} \Gamma_{em}$) although the nature of the
 measured turbulence is electromagnetic($\beta \approx 0.6 $). The turbulence has been 
established as a consequence of electron temperature gradient (ETG) driven modes. Experimental observations of phase angle between density ($ n_e $) and potential ($\phi $) fluctuations, $ \theta_{\tilde{n}_e, \tilde{\phi}} $ and electrostatic particle flux, $ \Gamma_{es} $  shows good agreement with the corresponding theoretical estimates for ETG turbulence.
\end{abstract}
\maketitle
\section{Introduction} 

Turbulent transport is a ubiquitous phenomenon prevalent in laboratory, space and astrophysical systems \cite{Althaus, Boqdan, Tynan_PPCF2009, J. Connor, Aschwanden}. Transport in general determines the mean profile and the confinement properties of the system. In magnetic confinement fusion research, for example in tokomaks, one of the most important and burning issue is improvement of energy and particle confinement time for controlled thermonuclear reaction. The confinement properties degrade because of high outward particle and heat flux which is observed to be several magnitudes higher than the classical or neoclassical flux \cite{Wei}. This anomalous flux is attributed to turbulent fluctuations due to various instabilities inherent in the system. Confined systems are naturally inhomogeneous which act as source of free energy to drive the system, unstable to slightest of the perturbations over a desperate range of scales, from electron, ion to system size scale.

While ultimately measurements in high temperature fusion plasmas in toroidal geometry must be undertaken, but it is desirable to have a hierarchy of experiments for comparison with the goal of isolating important physical effects in simplest possible geometry. Linear devices like Columbia Linear Machine (CLM) \cite{Wei}, Large Volume Plasma Device (LVPD) \cite{Mattoo_PRL}, and by Moon et al \cite{Moon} have taken initiative in addressing some of the physical issues which are difficult to study in high temperature, toroidal devices.

Large scale perturbations are easy to probe in tokamaks and tremendous progress has been made on ion temperature gradient driven micro-turbulent mode and MHD modes. Electron larmor radius scale fluctuations ($\rho_e \sim \mu m $)  ) in the range of $k_\perp \rho_e < 1$ due to electron temperature gradient (ETG) is hard to probe in tokamaks due to extremely small scale length, though some progress has been made  in National Spherical Torus Experiment (NSTX) \cite{Ren} and Tore Supra \cite{Horton_2004}.

On the other hand it is possible to scale up the size of the ETG mode in a simpler setting of straight magnetic field line geometry. Recently, ETG turbulence was established in the target region of LVPD \cite{Mattoo_PRL}. This device can be divided into three regions – source region where plasma is produced by filament heating, the filter region which provides a strong transverse magnetic field over a radial extent of 100 cm($\pm50cm$) to stop the energetic electrons emitted from the filament and the target region where plasma appears after diffusing through the filter magnetic field. Fluctuation and turbulent transport studies reported in this paper are done in the target region at distance of $100cm $ from the EEF location.

Both electrostatic and electromagnetic particle fluxes due to turbulent fluctuations are measured across the radius of the target chamber. It is observed that the turbulent particle flux is radially inward across the radius of the target plasma. The electromagnetic flux is observed to be several orders of magnitude less than the electrostatic particle flux. Hence the detailed investigation is provided on the electrostatic particle flux. The equilibrium electron density and temperature profiles are also measured which shows centrally peaked profiles. This is consistent with the observed particle pinch.  How? Unlike in tokamaks where the particle fuelling is done at the edge, the plasma source in LVPD comprises of heated tungsten filaments arranged in the periphery of a rectangle, coaxial to the device located at one axial end. Clearly, the filaments are not exactly at the edge, but a bit inside and not shodowed by the excited filter region. Since the axial diffusion is much larger than the transverse diffusion, the plasma formed at one axial end spreads rapidly and along the field lines and hence the effective particle source can be thought to be axially elongated and slightly inside the radial edge. Clearly, central density can build up only if there is an inward particle transport i.e., a particle pinch. 

In this paper, the electrostatic particle flux is measured, compared and a theoretical explanation is provided for the observed behaviour of it across the radius. The particle flux results due to the phase difference between the density and potential fluctuation, different from 180 degree. The cross phase angle obtained in experiment matches well with the cross angle due to the non-adiabatic ion response resulting from the perpendicular resonance of the ETG mode with the ions. The net fluctuation induced flux is found inward directed. The flux obtained by this model is some hybrid of pure diffusion and thermo-diffusion since it cannot be split in these two parts clearly. However in the flat density region the flux becomes purely thermo-diffusive. It is also found that the thermodynamic entropy of the system is reduced due to inward particle flux. This means that the heat flux, wave particle energy exchange, external sources and other dissipations together must overcome the entropy destruction by inward particle flux for net entropy production.

The rest of the paper is organized as follows; the experimental setup and diagnostics is described in section I. The experimental results are discussed in section II. In section III, a summary of experimental results are given and a discussion on comparison of experimental results with the theoretical estimation is provided in Section IV.
\section{Experimental setup and diagnostics}\label{sec:experiment}
\begin{figure}[ht]
\centering{\includegraphics[scale=0.27]{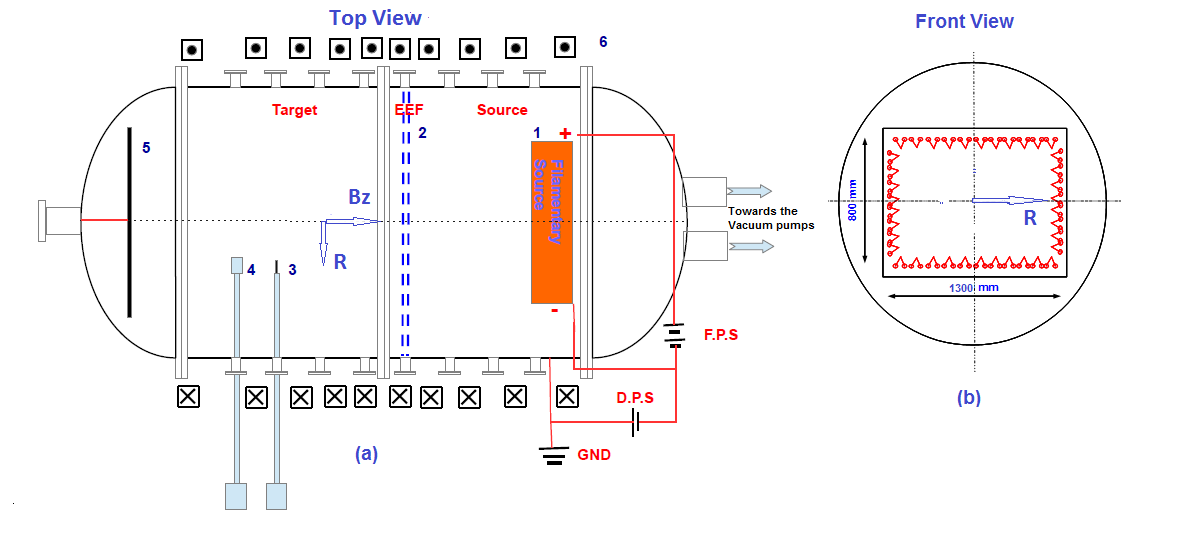}}
\caption{(a)Schematic diagram of Experimental setup, the layout of the internal component is marked as (1) back plate (2) EEF coil assembly (3) Langmuir Probe (4) a pair of B-dot and Langmuir probe , (5) end plate and (6) magnet coil system,  (b) cross sectional view of LVPD showing the  filament assembly arranged in rectangular geometry $1300mm\times800mm$ in the source region}
\label{fig:device} 
\end{figure}
\begin{figure}
\begin{center}
\includegraphics[scale=0.25]{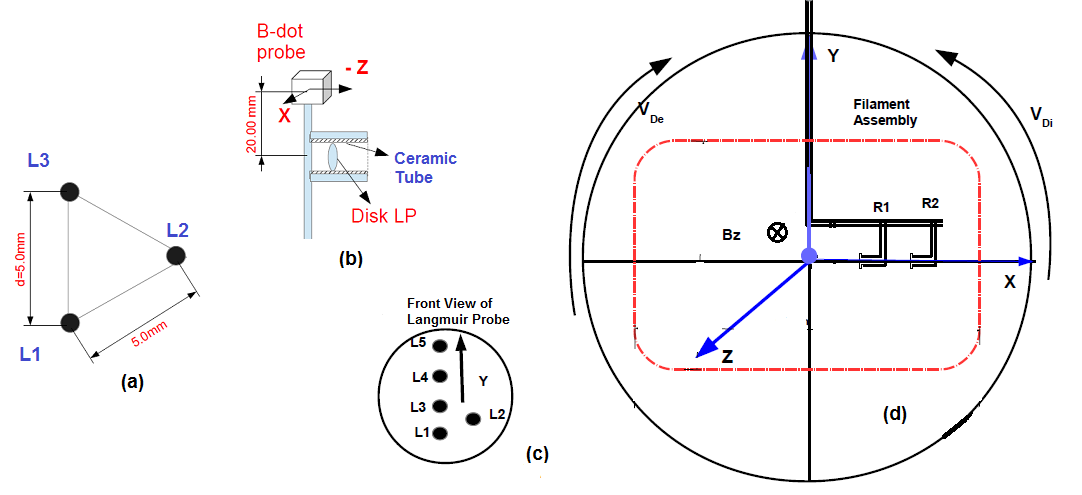}
\caption{Schematic of probe Assembly (a)Langmuir probe($l=8$ mm, $dia.=0.8 $ mm) arrangement in $\Delta$ configuration for electrostatic particle flux measurement (b)probe arrangement for simultaneous measurement of magnetic fluctuation($\tilde{B_r}$)and parallel current fluctuation for particle flux measurement. This contains a 3-axis B-dot probe along with specially arranged disk Langmuir probe with $ Dia = 5$ mm  (c)probe configuration for $k_\perp$ - $\omega$ measurement (d)Probe assembly from the top of the LVPD for simultaneous measurement of particle flux at two different radial location }
\label{fig:verB}
\end{center}
\end{figure}

The experimental setup consists of: (1)the LVPD \cite{Mattoo_RSI} device [FIG.~\ref{fig:device}] (2)the Electron Energy Filter(EEF) \cite{SKSingh_RSI}, (3) particle flux probes for diagnostics of electrostatic and electromagnetic fluctuations (4)the PXI based data acquisition system. The LVPD is a double walled, water cooled SS304 vacuum chamber having a diameter $ \sim 2m $  and length $ \sim 3 m $  supplemented by a combination of rotary-root-diffstak pumps, capable of pumping the system to a base pressure of $ 2 \times 10^{-6} mbar $.\\ The plasma source is a directly heated tungsten wire based multi- filamentary source $(2000K)$ with $ 36 $ number of hairpin shaped filaments ($ \varphi =0.5 mm, l=180 mm $) arranged on the periphery of a rectangular ($130cm \times 80cm$)back plate [figure \ref{fig:device} (b)]. Pulsed plasma (Argon gas, Pressure $4\times10^{-4}$ mbar and $\Delta t_{discharge}= 9.2ms $) is produced by applying a discharge voltage of $70$V between the plasma source and anode (vacuum vessel) in an ambient axial magnetic field, $Bz= 6.2$G, produced by
a set of $10$ coils, garlanded on LVPD.\\ The Electron Energy Filter (EEF) is a rectangular shaped with varying aspect ratio solenoid consisting of 155 numebr of individual coils arranged in $ 19 $ sets having a width of $ 4 cm $. Each set has been designed with equal resistive electrical path length. The EEF is coupled to a capacitor bank based pulse power suply capable of supplying a maximum current of $ 5 kA $ for pulse duration of $ 15 ms $ flat top \cite{5kA_RSI}. The EEF produces a uniform transverse magnetic field of $ \sim 160 G $ with input current of $ 2 kA $ for an EEF activation length of $ 1m $ that coprise of 13 identical central coils. EEF has divided the LVPD plasma into three distinct experimental regions namely, Source, EEF and target regions. The source region comprise of multi-filamentary source, the EEF region is the plasma volume enclosed by the solenoid itself and the target region is the region that receives the diffused plasma from the sources region through the EEF region. The fields within the cross-section of the EEF and along the LVPD axis are shown in FIG \ref{fig:verB}. A comparison of the measured field with that obtained through simulation exhibits good agreement. The magnetic field produced by EEF along the z-axis shows a rapid reduction in its value outside the EEF  boundary and attains level of $1$G at a distance of $20$ cm from the centre of the solenoid on either side of target and source plasma. This ensures the region of our study is unaffected by the EEF field.

\begin{figure}[ht]
\begin{center}
\includegraphics[scale=0.45]{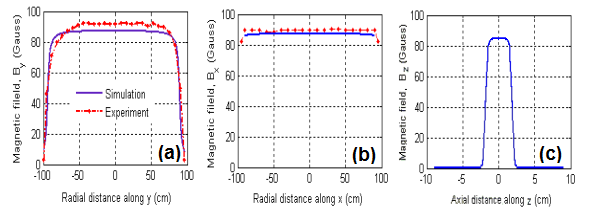}

\caption{The EEF produced magnetic field profile: (a) axial component, $ B_x $  of the EEF magnetic field along the axis of the 
solenoid, (b) radial component of the EEF magnetic field, $ B_y $ in the vertical(y) direction and (c) along the axis 
of the LVPD, i.e. $ B_z $ the across the axis of the EEF}
\label{fig:LP}
\end{center}
\end{figure}

The experiments are carried out in the target region of LVPD where ETG relevant plasma conditions are satisfied that includes finite electron temperature fluctuations ($\delta T_e$). Plasma parameters are measured by using Langmuir and B-dot probes. The plasma potential $\phi_p$ is measured by using a hot emissive probe \cite{smith_RSI}. The electron temperature, $T_e$ is determined from the I-V characteristic of the Langmuir probe. The azimuthal wavelength and phase velocity of mode are measured by an array of cylindrical Langmuir probes (N $ =4$, diameter $ = 0.5$ mm, length $ =8 $ mm, and separation, $ \Delta y=5 $ mm) mounted on a radially movable shaft. The fluctuation data and mean parameters are recorded at sampling rates of $500kSa/sec$ and subjected to bandpass filter with lower and upper frequency cut off of $300$Hz to $300$ kHz, respectively. Schematic of probe assembly is shown in figure \ref{fig:LP}.\\

The electrostatic particle flux data in the region is captured using a specially designed three probe assembly in $ '\Delta' $ configuration with length of vertices as, $ d=5 mm $. In this configuration, two probes separated vertically ($ L1 \& L2 $) are measuring the floating potential fluctuations ($ \phi_f $) and third probe ($ L2 $) intercepting different magnetic field line is used for measuring the density fluctuations ($ \delta n_e $) see figure ~\ref{fig:LP}(a). We have also configured probe assembly with two vertically separated emissive probes in order to measure the plasma potential fluctuation ($ \phi_p $) directly. This is carried out to see the effect of temperature fluctuation ($ \delta T_e $) over the measurement of poloidal electric field fluctuation ($ \delta E_{\theta} $) by the use of measured different potential fluctuation. We preferred the probe configuration with three Langmuir probes as our measurements showed that the fluctuation amplitudes for both floating and plasma potential measurement for the measurent of particle flux have no significant variation ($ 1 \sigma $). The tungsten wires ( $dia., \varphi =0.8mm$, and $ length, l =8mm $ ) are used for Langmuir probes construction. They are mounted on a radially moving linear probe drive with probe shaft capable of providing travel length of $ 1.2 $ meter within the vacuum chamber. The ion saturation current signal is obtained by biasing the probe at $ -80 V $ , and the floating potential measurements are carried out with unbiased floating probes terminated across a high impedance ( $ 1 M \Omega$). The large data length of more than $ 2 \times 10^{5} $ data points is used for power spectral analysis and is obtained from ensemble of approximately  identical plasma discharges. The data series is constructed by extracting $ 2048 $ data points from the steady state period of $ 6ms-8ms $  from each plasma discharge. This data is segmented into ($ 200 $) bins of  data points each for obtaining a higher frequency resolution in spectral analysis. 

In finite beta plasma ($\beta\sim 0.6$) conditions of target plasma, we carried out measurement of magnetic fluctuation induced particle flux ( $ \Gamma_{em} $) by simultaneous measurement of correlated fluctuations in $B_r$ and parallel streaming electron flux,$J_{|| e}$. The fluctuations in magnetic field are measured using a $ 3 $- axis magnetic probe (Bifilar configuration, Loop diameter = $10$ mm, turns, $ N = 30$, $L=10$mm). The pickup coil is calibrated using a known magnetic field of a Helmholtz coil. A transfer function for the pickup coil is used to derive the magnetic field from the pickup voltage, $V_{loop}$ . Since the output voltage of B-dot probe is proportional to its cross-sectional area ($A$), the number of turns in the coil($ N $), and the time characteristics of the magnetic field($\frac{dB}{dt}$). Hence the obtained output voltage from a B-dot prove is given by $V(t)= NA\frac{dB}{dt}$. So the time varying magnetic field can be obtained as $B(t)=\frac{1}{NA}\int_{}^{} V(t) dt$. Therefore to obtain the magnetic field signal we use RC-integrator with time constant($RC\sim$) $1$ ms such that $\frac{1}{RC}<\ \omega_{tur}$,where $\omega_{tur}$ is frequency of turbulence. The measured magnetic field is compared for both its amplitude and the frequency spectra. The parallel electron current is measured by a disc Langmuire probe (figure ~\ref{fig:LP}(b)) accommodated well inside a cylindrical metallic(SS304) tube with a isolation of ceramic tube.The parallel electron current is measured by a disc probe accommodated well inside a cylindrical hollow ceramic tube. The probe is placed in such a manner that it hardly intercepts the ion current and measures only electron contribution. The fluctuations in parallel electron current density, $ \delta J_{|| e}$   is measured by keeping the disc probe biased at plasma potential ($ \phi_p $ ). Both 3-axis probe assembly and the disc probe are mounted on a single ceramic mould so that it samples same plasma.

The data acquired for different parameters is captured in PXI based $ 40 $ channel fast data acquisition system. Out of total  $ 40 $ channels, $ 32 $ are single ended channels with maximum sampling rate of $ 60 MSa/s $ , digitization rate of $ 10 $ bit, and record length of $ 250 kpts $ and $ 8 $ single ended channels with  maximum sampling rate of $ 1.25 GSa/s $  , digitization rate of $ 12 $ bit and a record length of   $ 12 Mpts$. The data is retrieved to local computers for post processing.

\section{Experimental Observation}\label{Observation} 

\subsection{Temporal Evolution of Plasma}
\begin{figure}[ht]
\centering{\includegraphics[scale=0.25]{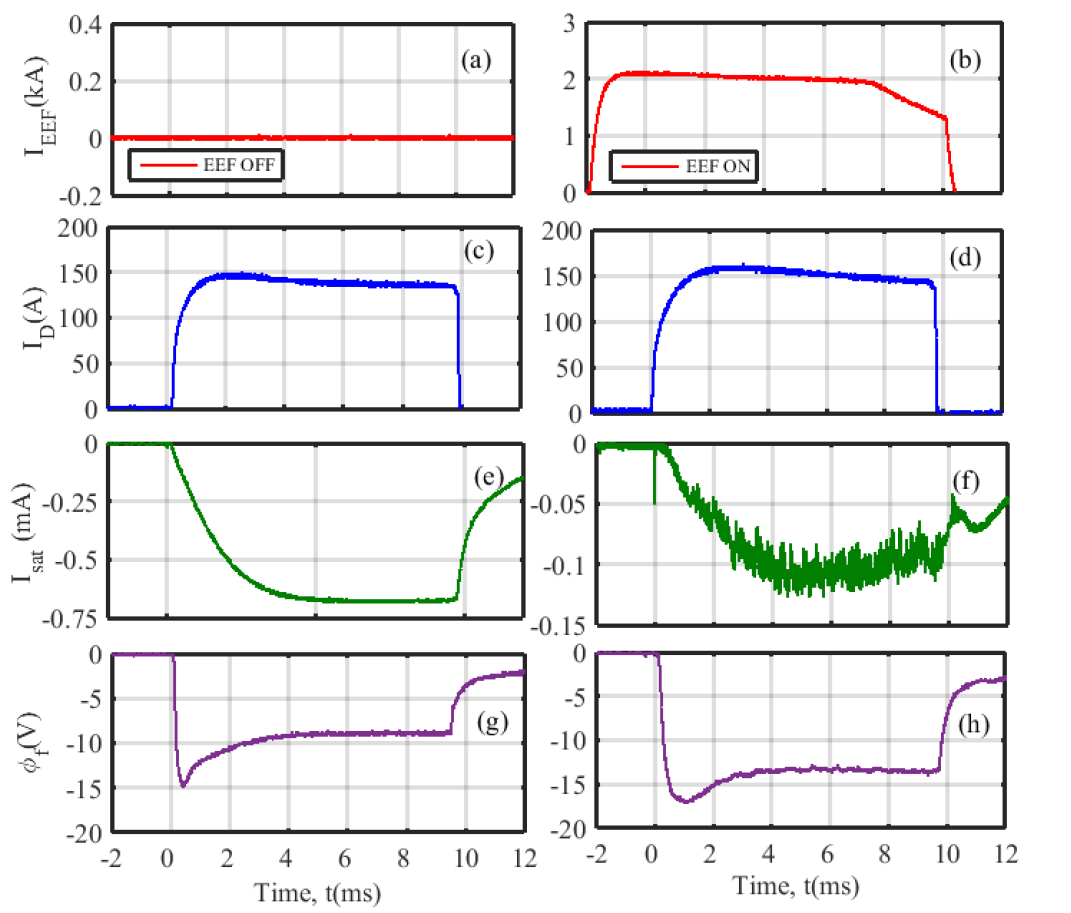}}
\caption{Traces each of, (a-b) EEF filter current, $I_{EEF}$,(c-d) Discharge Current, $I_d$, (e-f) Ion-saturation current, 
$I_{sat}$, (g-h) Floting Potential, $\phi_f$. The ion saturation current, which mimics the plasma density, 
exhibits fluctuation for plasma density, $n_e \sim 5\times 10^{16} m^{-3}$. The traces are obtained at 
$x=0$, $z=-100cm$ from the EEF. }
\label{fig:discharge} 
\end{figure}

To understand the turbulence features in plasma, the basic plasma parameters such as electron temperature, $T_e$, plasma density, $n_e$, and floating potential, $\phi_f$, and there fluctuations are measured by using conventional Langmuire probe within the plasma pulse duration. The typical temporal profiles of pulsed plasma parameters in the target region in the presence of EEF OFF and EEF ON are shown in figure~\ref{fig:discharge}.

\begin{figure}[ht]
\begin{center}
\includegraphics[scale=0.25]{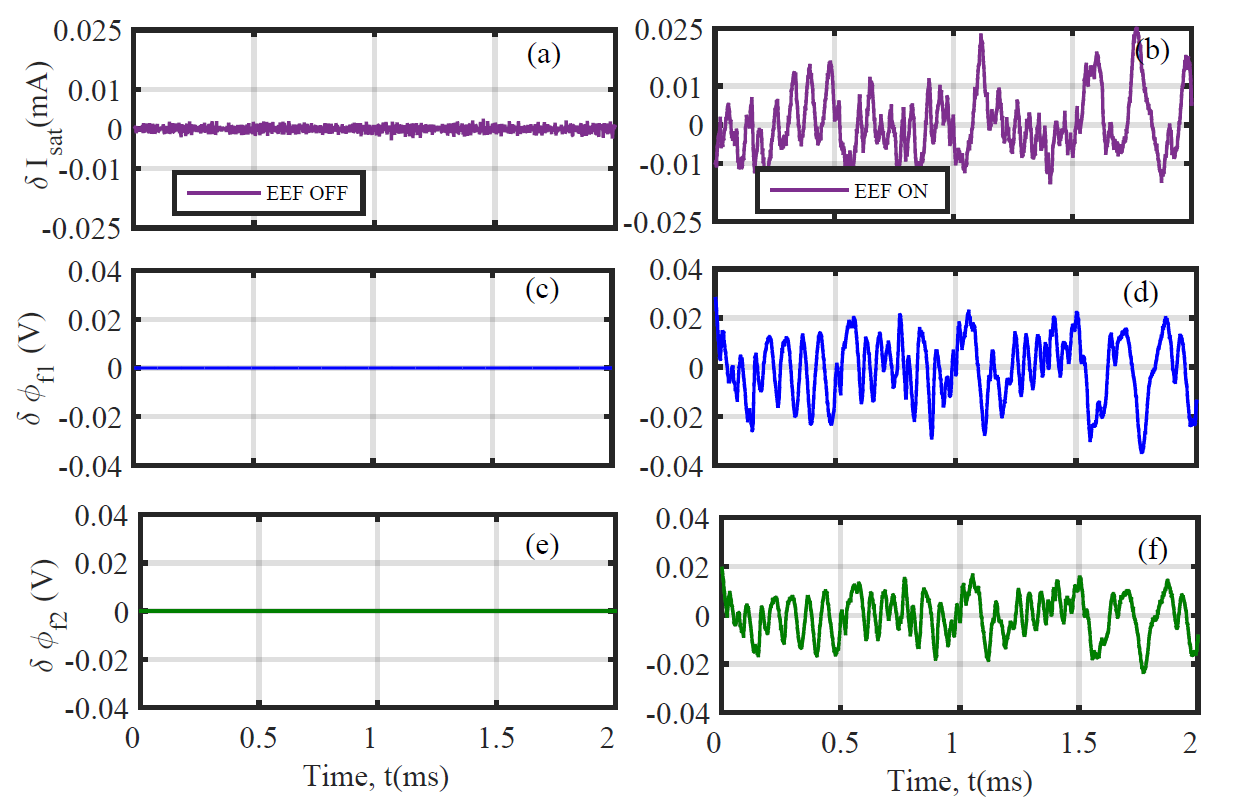}
\caption{Fluctuation Time Series of $\delta I_{sat} $, $\delta \phi_{f1}$ and $\delta \phi_{f2}$ obtained from Langmuire  probes for EEF active(ON) and inactive(OFF) cases are presented in (a)-(b),(c)-(d) and (e)-(f) respectively. Fluctuation exists only when ETG scale lengths satisfied by plasma profile}
\label{fig:fluctuations}
\end{center}
\end{figure}

In the experiments, the plasma discharge pulse is accommodated within the EEF pulse length. Figure~\ref{fig:discharge} shows the plasma characteristics measured in the traget region for EEF OFF and EEF ON cases. Figure~\ref{fig:discharge}(a)-\ref{fig:discharge}(b) shows the typical EEF dischage pulse. Figure~\ref{fig:discharge}(c)-\ref{fig:discharge}(d) shows the typical discharge current profile with EEF OFF and EEF ON cases in LVPD. Figure~\ref{fig:discharge}(e)-\ref{fig:discharge}(f) shows the ion saturation current ($ I_{sat} $) attains steady in the pulse duration of $ 4ms \leq t <9ms $ for both the EEF OFF and EEF ON conditions but it is evident from this figure that the fluctuation levels is more prominent in EEF ON case than EEF OFF case. Since the ion saturation current, ( $ I_{sat} $ ), measurement mimics the plasma density, in this view we can assure that the density level can be reduced by almost a factor of $ 8 $ in EEF ON case than EEF OFF case. Figure ~\ref{fig:discharge}(g)-\ref{fig:discharge}(h) shows the time evolution of floating potential for EEF OFF and ON cases. The mean floating potential shows an increase with EEF ON case.\\
 A typical time profile of  ion-saturation fluctuation, ( $ \delta I_{sat} $ )and potential fluctuations ( $ \phi_{f1} $ and $ \phi_{f2} $) in the core region of target plasma ( $ R=30 cm$  ) is obtained using flux probe assembly for EEF OFF and EEF ON cases as shown in figure ~\ref{fig:fluctuations}. We observed that fluctuations are significantly high when EEF is ON and the fluctuation levels remain close to the noise level when EEF is OFF.  


\subsection{Radial plasma profiles}\label{sec:plasma charecteristic}
\begin{figure}[ht]
\centering{\includegraphics[scale=0.35]{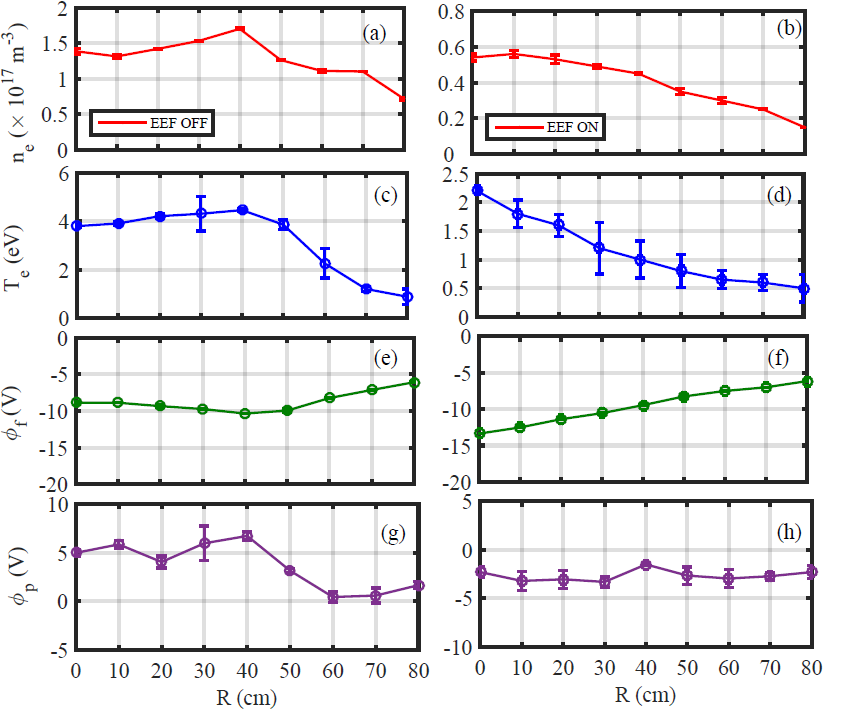}}
\caption{Radial profiles of mean parameters,(a)-(b) plasma density, $ n_e $, (c)-(d) electron temperature, $ T_e $,(e)-(f) floating potential profile,$ \phi_f $, and (g)-(h) plasma potential profile, $ \phi_p $ for EEF ON and EEF OFF, respectively. It is observed that the finite ETG of scale length, $L_{T_e}\sim 50 cm$ exists only in the core region of the LVPD plasma for EEF ON condition.}
\label{fig:mean_profiles} 
\end{figure}

Mean radial profiles of basic plasma parameters are shown in figure \ref{fig:mean_profiles}. The plasma density, $ n_e $, electron temperature, $ T_e $, floating potential $ \phi_f $ and plasma potential $ \phi_p $, are measured for investigating the plasma for EEF OFF and EEF ON case. We carried out investigations for confirming performance of EEF for producing a Maxwellian plasma by validating the relationship $ \phi_p = \phi_f + 5.4 T_e $ in the target region of LVPD plasma. A detailed characterization of the plasma in the presence of EEF was reported by Sushil \textit{et al}\cite{SKSingh_PPCF}, where it was shown that various scale lengths of density ($ n_e $) and electron temperature($ T_e $) are possible by suitably configuring various elements of EEF.

\begin{figure}[ht]
\centering{\includegraphics[scale=0.30]{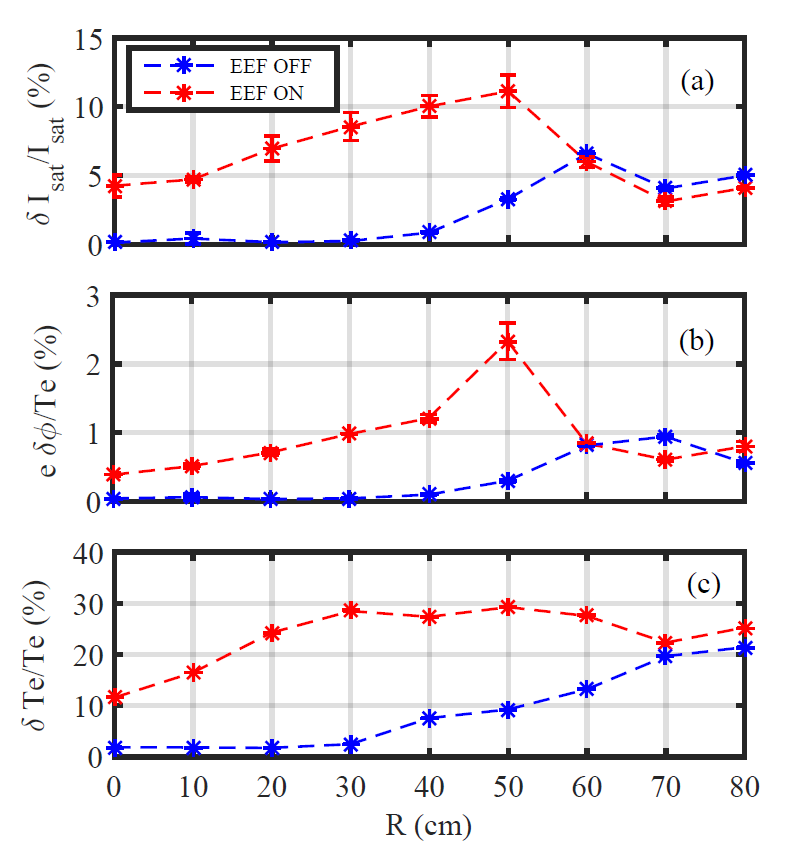}}
\caption{Radial profile showing comparison of fluctuation in a) density, $ \tilde{n}_e $($ = \delta n_e/n_e$), b) potential fluctuation, $ \tilde{\phi}$ ($ = e\delta \phi_f/T_e $) and c) electron temperature fluctuation, $\tilde{T}_e $ ($ = \delta T_e/T_e $) for EEF OFF (blue color) and EEF ON(red color)}
\label{fig:fluc_radial}
\end{figure}

Although, various configurations of the EEF are excited, we restricted our measurements here to the best two cases when ETG conditions is not satisfied i.e. EEF OFF case and when ETG condition is satisfied i.e. EEF ON case in the target plasma of LVPD for studying equilibrium profiles and fluctuations of plasma parameters. For EEF OFF case, the profile of plasma density, $ n_e $, and electron temperature, $ T_e $, is flat in the core region ($ R \leq 50 cm$). The floating potential, ($ \phi_f $) and the plasma potential, ($ \phi_p $), profiles also follow the density and temperature profiles in the core of target plasma . These profiles shows finite gradient in the outer region ($ R > 50$). The profile flatness in the core plasma for EEF OFF case does not satisfy the ETG threshold conditions. As a consequence no significant fluctuations are observed in the core plasma when EEF is OFF [figure ~\ref{fig:fluctuations} ((a), (c) and (e))]. For EEF ON case, the core plasma exhibits a flat density profile ($ L_{n_e} \sim 600 cm $) but sharp gradient in electron temperature profile ($ L_{T_e} \sim 50 cm $). The radial floating potential ($ \phi_f $) profile shows a gradient but the plasma potential profile remains flat. This assures the absence of radial electric field in the core plasma. The gradient scale lengths of density, $ L_{n_e}={d(ln n_e}/{dr} $, and electron temperature, $ L_{T_e} = {d(ln T_e)}/{dr} $ satisfying the ETG turbulence threshold conditions ($ \eta_e = L_{n_e} / L_{T_e} > 2/3 $). Existence of ETG turbulence is seen as significant enhancement of density and potential fluctuations in this region [figure ~\ref{fig:fluctuations}((b), (d) and (f))]. The core region is dominated solely by the gradient in electron temperature whereas the outer region has pressure gradients. There is no electric field present resulting no $ E \times B $ rotation in ETG dominant region.The core region has both electrons and ions well confined by the applied magnetic field ( $ B_z \approx 6.2 G $)but as ion larmor radius, $\rho_i \sim 45$ cm, the plasma density gets flattened on that scale. The electron temperature follows the electron larmor radius, $\rho_e $ as it exhibits a significant gradient. The turbulence in the core region is dominated by electron temperature gradient but in outer region, it is primarily due to the pressure gradients. The ions remains unconfined in LVPD target region and they have tendency to move radially out in successive collisions. Electrons follow the field lines and move primarily along the axis in order to maintain quasi-neutrality. 
The radial profiles for density ($ n_e $ )and potential ( $\phi_f$), temperature ( $ T_e $ )  fluctuations are investigated in the core and edge regions and compared for EEF OFF and ON cases. Experimentally special attention is given to the core turbulence where ETG is well established. We observed that fluctuation amplitudes are higher in the core region when EEF is ON. The typical level of normalized fluctuations obtained for density fluctuation, $ \delta n_e / n_e $, potential fluctuation, $ \delta \phi_f / T_e $ and temperature fluctuation, $ \delta T_e / T_e $  are $ 5 \% -10\%$, $ 0.5 \%- 2.5 \%$  and $ 10\%-30\%$   respectively [figure ~\ref{fig:fluc_radial}]. These fluctuations in the core region approach nearly noise level when EEF is OFF. In edge region for EEF OFF case, the level of fluctuations increases, reason for this may be the enhanced gradient observed in the mean plasma profiles and wall effects.  It is interesting to note here that the temperature fluctuations remain low in EEF OFF case. Also, in the edge region, electron temperature gradient is insignificant for EEF ON plasma. 
\\

\subsection{Cross Correlation}
\begin{figure}
\begin{center}
\includegraphics[scale=0.22]{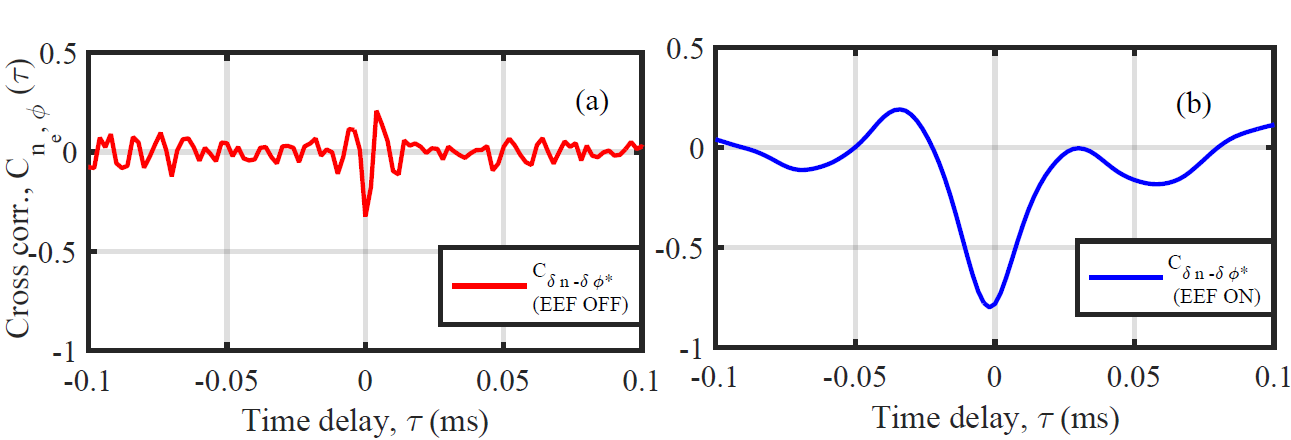}
\caption{ The cross correlation coefficient for density and potential fluctuations measured in the core region ( $ R = 20 cm $) for EEF OFF and EEF ON cases.}
\label{fig:Cnphi}
\end{center}
\end{figure}

Nature of turbulence and their mutual correlation can be understood better by studying cross-correlation,
and power spectra. These are necessary for identifying the nature of instability.
We have measured the correlation coefficients between density and potential fluctuations for EEF OFF and ON cases as shown in Fig ~\ref{fig:Cnphi}.

The normalized density fluctuations $\delta n_e /n_e $ and the potential $e\phi_f/kT_e$ are found to be strongly anti-correlated which is for ETG, $ \tilde{n} \approx - \tau^{*} \tilde{\phi}$. 
The correlation coefficient $C_{\delta n \delta \phi} \approx -0.8$. Its value reduces
to $-0.2$ and becomes weakly correlated when EEF is switched off. These measurements are
carried out using array of three probe assembly, where probe separation between the pair
probes is $5$mm. There may be a slight spatial de-correlation as the probes used are not located
on the same magnetic field line.\\
\subsection{Correlation Length}
\begin{figure}[h]
\begin{center}
\includegraphics[scale=0.25]{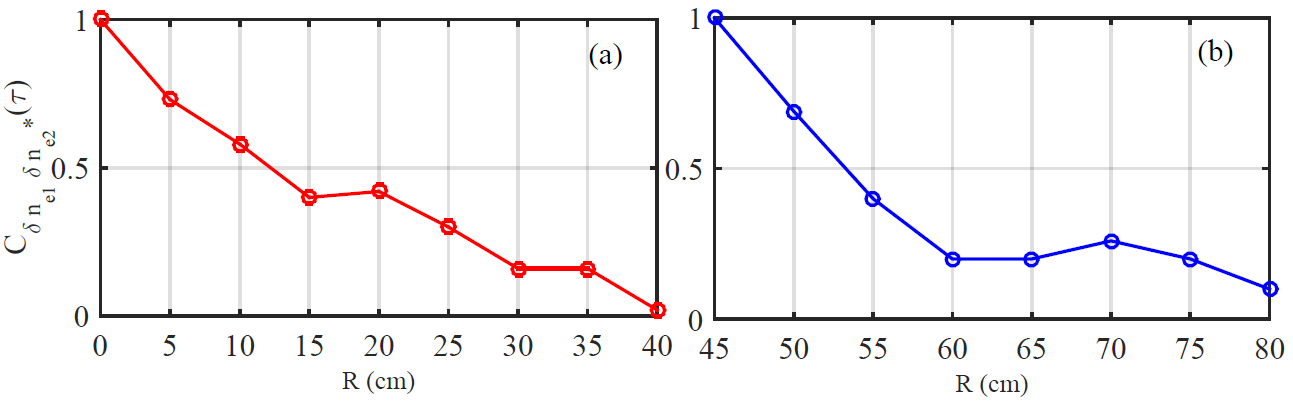}
\caption{Radial correlation plot obtained in the (a)Core Region (b) Edge Region}
\label{fig:radialcorr}
\end{center}

\end{figure}

The radial correlation lengths are obtained for the density fluctuations in the ETG dominated
and edge regions. For this, a pair of probes is used and one of the probe is kept stationary
while other moves radially outward in a step of $ 5 $cm. These measurements are undertaken so
as to establish the region where observed fluctuations remains correlated. In ETG dominated
region, this distance comes out to be $\sim 25$ cm and in the edge region, it reduces to $ \sim 12 $ cm
respectively as shown in figure \ref{fig:radialcorr}.\\

\subsection{Power Spectra Plot}
\begin{figure}[ht]
\begin{center}
\includegraphics[scale=0.25]{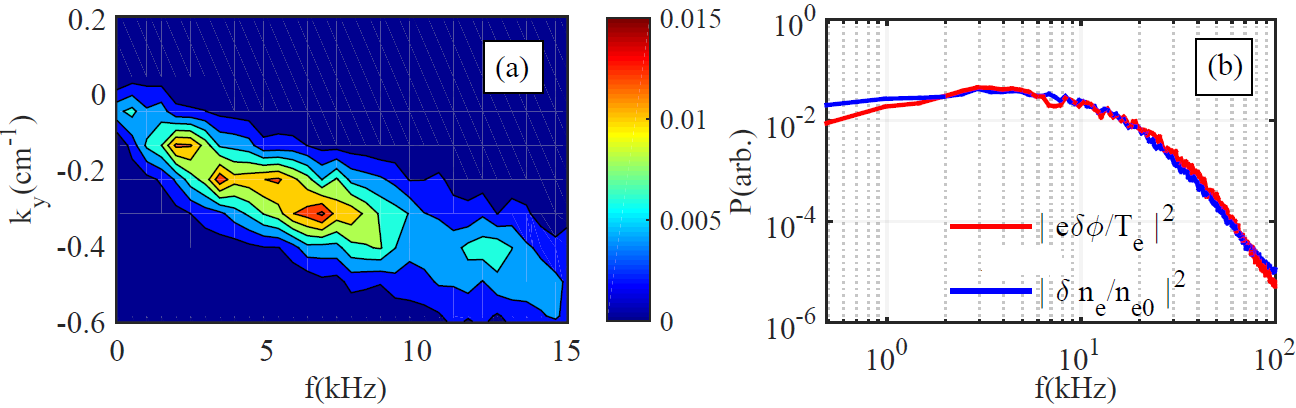}
\caption{(a) The joint wave number-frequency spectra for density fluctuations in sub-plot (a). The mode exhibits peak power at $f\sim 4 kHz$ and sub-plot (b) shows the broad band nature of turbulence with significant power residing within $f<50kHz$ for both density ($n_e$) and potential ($\phi$) fluctuation at $R=30 cm$ at the core of Target plasma of LVPD }
\label{fig:skw_R30}
\end{center}
\end{figure}

The turbulence observed in the core region exhibits broad band spectra
with significant power between $1-15$ kHz. As shown in figure \ref{fig:skw_R30}(a) the mode frequency observed in
the lower  hybrid range, i.e. $ \omega_{ci} < \omega < \omega_{ce} $, where $ \omega_{ci}$, ion cyclotron frequency, $ \omega_{ce} $, electron cyclotron frequency also satisfying the following characteristics such as $ k_\perp \rho_e \leq 1 $, $ k_\perp \rho_i \gg 1 $, and $ k_\perp c_{i} \gg \omega $, where $ \rho_e, \rho_i $ are electron Larmor radii's respectively, which suggest that the instability driving the turbulence is ETG. The joint wave number–frequency spectrum $ S(k,\omega) $ is determined for the $ n_e $ fluctuations as shown in figure \ref{fig:skw_R30}(a). We have used data obtained from probes separated in the vertical direction with probe spacing $ 0.5 $ cm. The spectrum peaks at  $ (\omega/2\pi), f \sim 4 $kHz with perpendicular wave number $k_{\perp} \sim 0.15$ $cm^-1$. The spectrum exhibits a width in frequency, $ \delta f/f \sim 1.8 $, and wave vector $ \delta k / k \sim 2.2 $, supporting its broad band nature. The plasma fluctuation has a long poloidal wavelength, $ \sim 40 $ cm. The phase velocity of the observed mode is , $v_{ph} \approx 10^5 cm/s $ in poloidal direction. We have put probe identifiers for the correct assessment of drift direction (FIG \ref{fig:LP} ). The k-spectra shows an asymmetrical distribution of power to different modes.
The auto power spectra of density ($n_e$) and potential ($\phi$) fluctuation  presents a broad band spectra with peak power residing within frequency, $f < 50$ kHz [figure \ref{fig:skw_R30}(b)].

\subsection{Phase Velocity}
\begin{figure}[h]
\begin{center}
\includegraphics[scale=0.25]{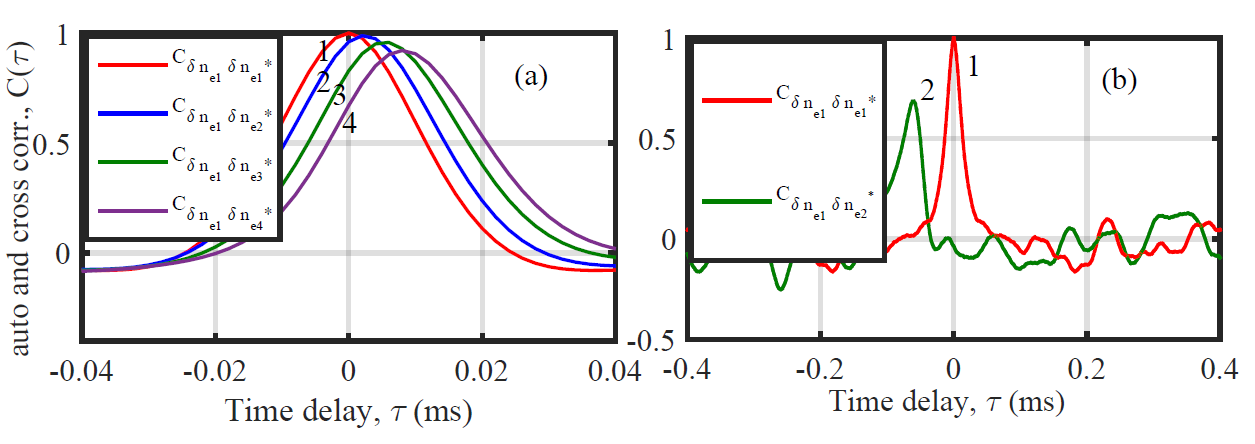}

\caption{The auto-correlation plot of density fluctuation measurement $\delta n_{e1}$ at $R=20 cm$ (red color) and cross -correlation of density fluctuation  $\delta n_{e2}$ at $R=30 cm$ with $\delta n_{e1}$ (blue color) for radial propagation measurement}
\label{fig:phase_velocity}
\end{center}
\end{figure}

The radial phase velocity for density fluctuations shows that the fluctuations propagating radially inward towards the core of the device. The measurements are carried out by using two radially separated Langmuir probes kept at $ R = 20 $ and  $ R= 30 cm $ respectively. Fig [\ref{fig:phase_velocity}] shows the auto and cross correlation for density fluctuations measured by the probes. From the cross correlation time, the radial phase velocity is measured to be, $ V_R \approx -1.5 \times 10^{5} cm/sec $. Same measurements were repeated at different radial locations in the core plasma and are found that the fluctuation phase velocity is radially inward. 
Similarly, poloidal phase velocity has been measured by using two vertically separated Langmuir probes ($ \Delta y= 5 mm $ ). The measurement shows that the order of poloidal phase velocity ( $ V_{\theta }$ ) is comparable to radial phase velocity of the fluctuations. 

\section{Plasma Transport}

In this section we will show the measurement that will estimate the particle transport due to the ETG driven fluctuations in the core target plasma of LVPD and hence the net plasma transport. We will calculate the fluctuation induced electrostatic flux and electromagnetic flux in order to estimate net particle flux due to fluctuation and in the next section we will compare the experimentally measured electrostatic particle flux with theoretical estimates.

\subsection{Electrostatic Particle Flux estimation}
\begin{figure}[ht]
\begin{center}
\includegraphics[scale=0.15]{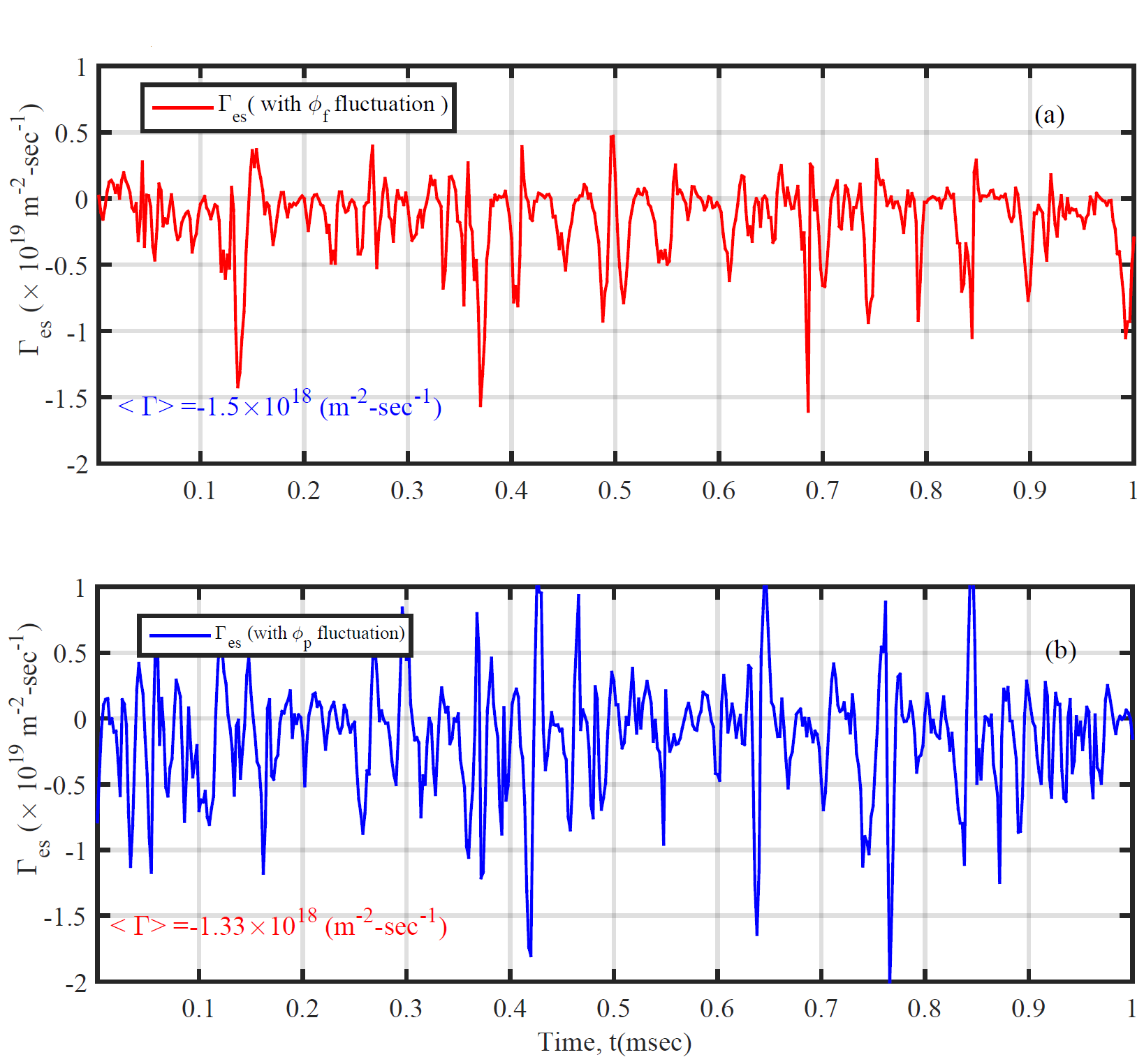}	

\caption{The time series of electrosatic particle flux measured at ETG dominated region($R=20$cm) by use of (a) Simple Langmuire probe (b) Heatable Langmuire probe (Emissive Langmuire probe)}
\label{fig:emissive_flux}
\end{center}
\end{figure}

\begin{figure}[ht]
\begin{center}
\includegraphics[scale=0.25]{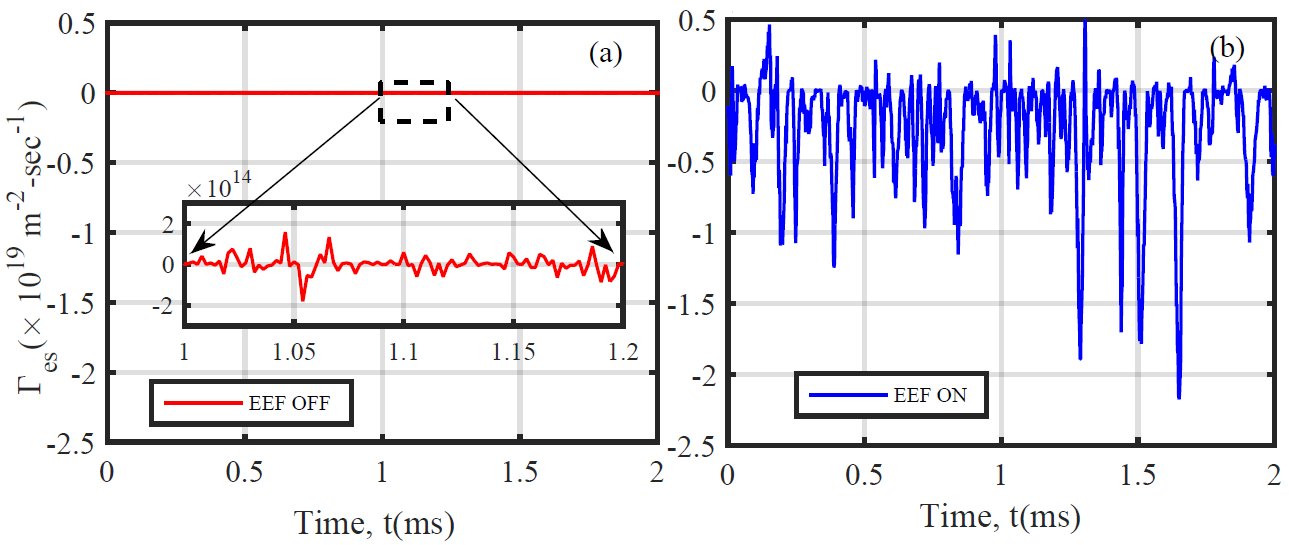}

\caption{The time series of electrostatic particle flux($\Gamma_{es}$)plot at R=30 cm for two cases (a)EEF ON (b) EEF OFF }
\label{fig:flux_time}
\end{center}
\end{figure}

The electrostatic particle flux ( $ \Gamma_{es} $) is measured from the correlated density fluctuation ($ \delta n_e $) and fluctuation radial velocity ($ \delta v_r $) estimated from poloidally fluctuating electric field ($ \delta E_{\theta} $) by $ E \times B $ drift, where $ E_{\theta} $ is measured by two poloidally separted probes as $ E_{\theta} = - (\delta \phi_{f2}-\delta \phi_{f1})/d $, where $ d $ is probe separation. Then the electrostatic particle flux is calculated as, $\Gamma_{es} = < \tilde{n}_e \tilde{V}_r > $. In the temperature fluctuations, electric field fluctuation measurement for estimation the particle flux with floating potential fluctuation measurements may not yield correct results. We therefore have compared the electrostatic particle fluxes by the estimation of electric field calculated from the floating potential ($ \phi_f $) as well as plasma potential ($ \phi_p $). From fig \ref{fig:emissive_flux} it is clear that both measurement of particle flux are qualitatively and quantitatively are within the error bar. As there is no significant difference is observed in the flux estimation therefore, for our convenience we carried out further estimations of particle flux using floating potential techniques.
Further experiments are carried out for flux measured by Langmuir probes at R = 30cm for EEF OFF and ON cases. It is evident from the figure \ref{fig:flux_time} that the electrostatic flux is significantly enhanced ($ < \Gamma_{es} \approx -1.78\times 10^{18} m^{-2}-s^{-1} >$) when EEF is ON and ETG conditions are established. In the case of EEF OFF the electrostatic flux ($ < \Gamma_{es} > \approx $) remains insignificant and is lower by an order of ($ 10^{9} $)  from EEF ON case.  Also, it can be commented that the observed electrostatic flux in ETG region has significant and its negative sign suggests that the net particle flux induced due to the fluctuations are moving radially inward.

\begin{figure}[hb]
\begin{center}
\includegraphics[scale=0.25]{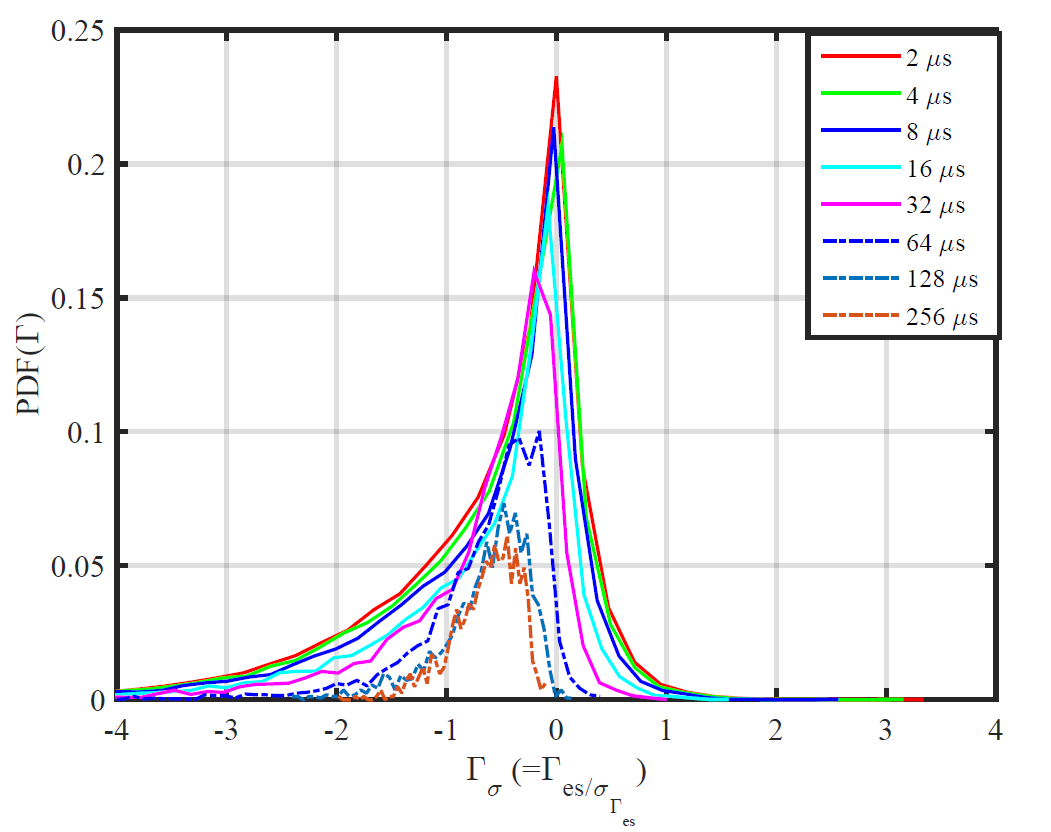}

\caption{The PDFs in the units of standard deviation ($ \sigma_{\Gamma_{es}} \approx 4.0 \times 10^{18} m^{-2}-s^{-1}$ for different averaging time, $\tau$ (see legend) of flux data}
\label{fig:pdf}
\end{center}
\end{figure}

To ensure the nature of particle flux in ETG region, the statistical analysis is performed using sufficient length of data points. The PDF analysis of particle flux is non-Gaussian in nature [figure \ref{fig:pdf}]. It is asymmetric and negatively skewed. The skewness and kurtosis obtained are $-1.62$ and $6.2$, respectively. The negative skewness indicates a predominance of large negative flux.\\

Following the work of Carreras \cite{Carreras}, the PDF for the fluctuation-induced turbulent flux $ \Gamma = \tilde{n}\tilde{v}_r $ is
\begin{equation}
P(\Gamma)= \frac{1}{\pi}\frac{\sqrt{1-\gamma^{2}}}{W_n W_{v_r}}K_{o}\big ( \frac{|\Gamma|}{W_n W_{v_r}} \big ) \\
\exp \big ( -\gamma \frac{\Gamma}{W_n W_{v_r}}\big )
\label{eq:eq_PDF}
\end{equation}
where $ \gamma $ measure the strength and sign of the correlation between density and velocity fluctuation $(|\gamma|<1) $, the parameters $W_n$ and $ W_{v_r} $ are the square of the variance of $\tilde{n} $ and $ \tilde v_r $, respectively. $ K_{o} $ is modified Bessel function and its argument is symmetric with respect to the flux direction. By the use of equation \ref{eq:eq_PDF} the averaged flux is derived as
\begin{equation}
<\Gamma>=-\frac{\gamma}{1-\gamma^2} W_n W_{v_r}
\label{eq:eq_meanflux}
\end{equation}

For the averaged flux to be outward, $\gamma < 0 $ and it is also an indirect measurement of relative phase between $\tilde n $ and $\tilde{v}_r $ and is described by the expression
\begin{equation}
\cos{\theta}=\frac{<\tilde{n}\tilde{v}_r>}{<\tilde{n}>^{1/2} < \tilde{v}_r>^{1/2}}=-\gamma
\label{eq:phase_eq}
\end{equation}

Taking the phase information from experimental observation as $\approx -130 ^{\circ}$, and the use of equation \ref{eq:phase_eq} the value of $\gamma = 0.70 $ which is greater than $ 0 $, suggesting averaged flux to be inward.
The calculated flux after measuring $W_n$, $W_{v_r}$ by the use of $ \gamma $ comes out $-7.23\times 10^{18} m^{-2}-s^{-1} $, having same order what we have obtained experimentally.
 
\begin{figure}[h]
\begin{center}
\includegraphics[scale=0.25]{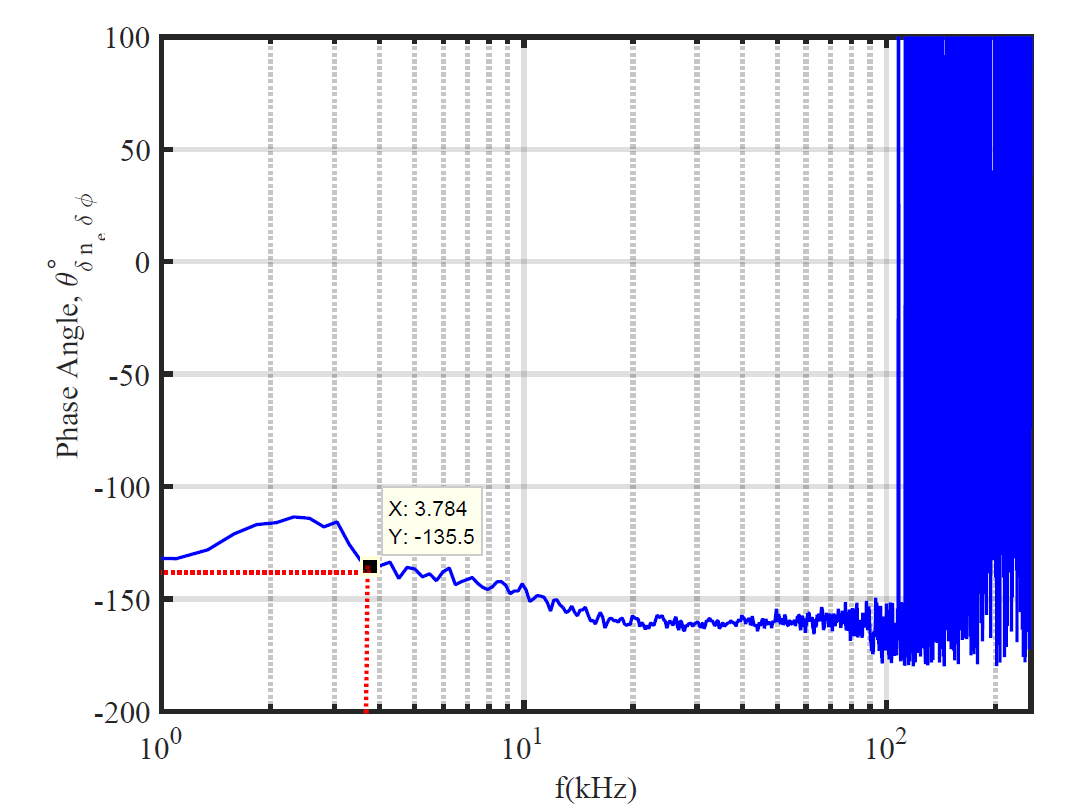}

\caption{Phase Angle plot between $\delta n_{e}$ and $\delta \phi$ at R=30 cm for ETG condition. The perpendicular line on both axes shows the phase angle value corresponding to maximum powered frequency at same location}
\label{fig:phase_plot}
\end{center}
\end{figure}

We have observed fluctuation induced finite inward particle flux from the correlated density and potential fluctuations. Theory predicts a finite phase delay ($ \theta_{\tilde{n}-\tilde{\phi}} $ ) between the density and potential fluctuations as a consequence of which a net turbulent particle transport takes place. We have measured the cross phase between the density and potential fluctuations as a function of frequency at $ R=30cm $[Figure \ref{fig:phase_plot}]. It can be seen that the cross phase angle corresponding to our frequency band of interest is finite and is significantly deviated from $ 180^{\circ} $ . It should be noted that for EEF OFF case, potential fluctuations are insignificant and are not correlated with density fluctuations hence the particle flux is negligible comparing with EEF ON case. 

\subsection{Electromagnetic Particle Flux}
\begin{figure}[ht]
\begin{center}
\includegraphics[scale=0.20]{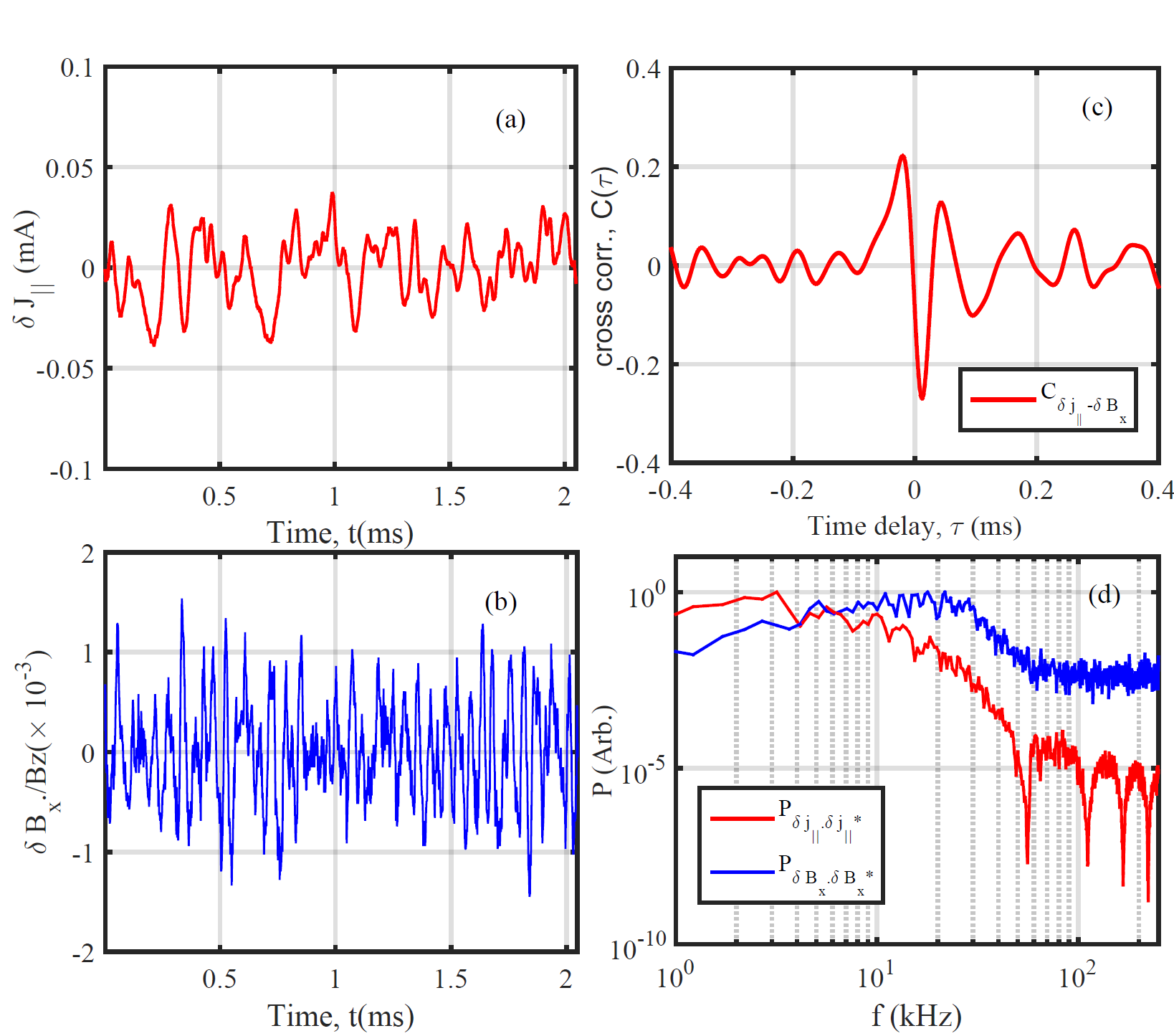}
\caption{(a)Time profile of normalized $J_{||}$ fluctuation (b) Time profile of normalized $B_r$ fluctuation (c) Cross correlation between $J_{||}$ and $B_r$ fluctuation (d) power spectra of correlated $J_{||}$ and $B_r$ fluctuation }
\label{fig:coh_power}
\end{center}
\end{figure}

\begin{figure}[ht]
\begin{center}
\includegraphics[scale=0.95]{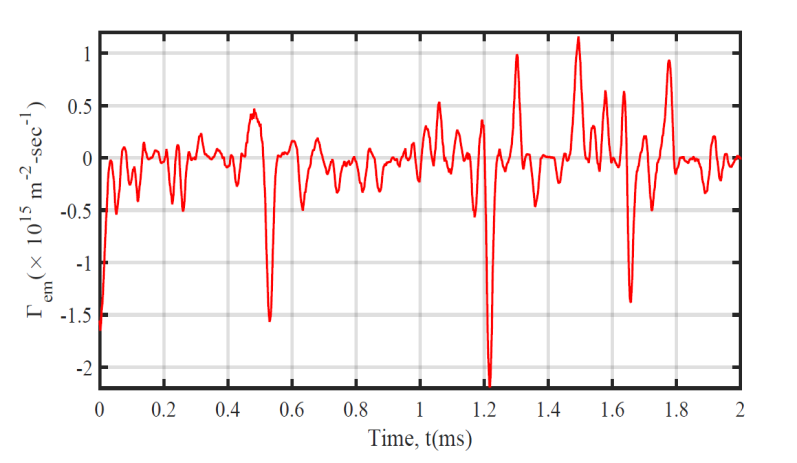}

\caption{Particle flux due to magnetic field fluctuation at $ R = 30 cm $}
\label{fig:emflux}
\end{center}
\end{figure}

To understand the electromagnetic effect of ETG turbulence over the particle transport, a experimental work  is carried out by  simultaneous measurement of correlated fluctuation in radial magnetic field ($B_r$) and parallel streaming electron flux,($J_{||}$). FIG \ref{fig:coh_power} shows the time profile of normalized parallel current fluctuation $\tilde{J_{||}}$ ($={\delta J_{||}}/{J_{||}}$) and radial magnetic field fluctuation $\tilde{B_r}$ ($={\delta B_r}/{B_z}$) with a level of $ 2.71 \% $ and $ 0.04 \% $ respectively. Their corresponding cross-correlation $C(\tilde{J_{||}}-\tilde{B_r})$ and power spectra are obtained which is shown in figure \ref{fig:coh_power} (c) and figure \ref{fig:coh_power} (d), respectively. From these plots we can infer that parallel electron current and radially magnetic field fluctuation share a common frequency band but they are weakly correlated as depicted from the cross correlation plot.  
The electromagnetic electron particle flux ($\Gamma_{em} $)can be estimated as $$ \Gamma_{em} = - \frac{<\delta J_{||,e} \delta B_r>}{e B_z}$$.
The particle flux obtained from the simultaneous measurement of both parallel electron current fluctuation,$\delta J_{||,e}$   and radial magnetic field fluctuation,$ \delta B_r $   is shown in figure \ref{fig:emflux}. The magnitude of obtained flux is of the order of $10^{11} m^{-2} s^{-1} $. On comparison with electrostatic counterpart, we observed that $<\Gamma_{em}>\approx 10^{-7} <\Gamma_{es} >$ , hence contribution of magnetic flux to total flux is negligible for this experimental observation.
The net fluctuation induced particle flux is sum of electrostatic and electromagnetic components, 
\begin{eqnarray*}
\Gamma = \Gamma_{es} + \Gamma_{em} = <\delta n \delta v_r > - \frac{<\delta J_{\parallel e} \delta B_r>}{eB}
\end{eqnarray*}

Irrespective of high beta plasma in LVPD, observation suggests that it’s the electro static counterpart only, which dominates the net turbulent particle flux.

\section{Theoretical Explanation}
\label{sec:theory}
\begin{figure}[h]
\centering{\includegraphics[scale=0.50]{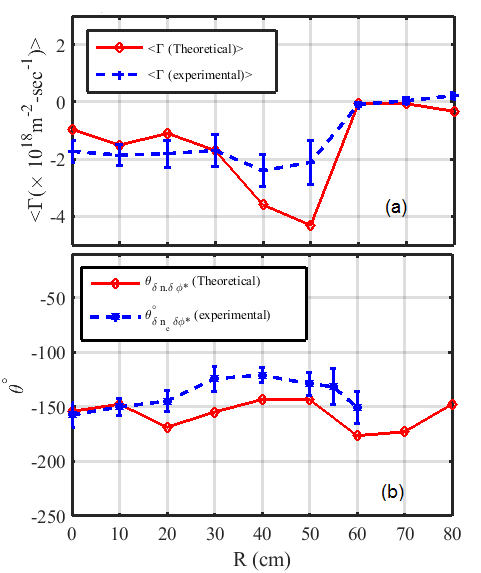}}
\caption{A comparison plot between experimental observed and analytically predicted values of (a) electrostatic particle flux,$<\Gamma_{es}>$ and (b) cross-phase angle, $\theta_{n_e-\phi}$ for ETG region in LVPD}
\label{fig:flux_phase}
\end{figure}

The radial profile of time averaged particle flux ($\Gamma$) and density($\tilde{n}$) potential ($\tilde{\phi}$) 
cross correlation angle ($\theta_{n\phi}$) is shown in Fig~\ref{fig:flux_phase} It is seen that the particle flux 
($\Gamma_{nv_{r}}$) is radially inward in the inner region and is roughly maximizing at radial location where $\eta_e$ 
is maximum. Also the cross angle is most deviated from $180^{\circ}$ where $\eta_e$ is maximum. The profile of particle 
flux is set by the product of density and potential fluctuation amplitudes and sinus of the cross angle. \\ Experimental 
results are compared with the theoretical model proposed by R Singh \textit{et al.}\cite{RSingh_2013} for turbulent flux in the electrostatic ETG turbulence. To make things clear we provide in the following the essential of model. Ions are considered as unmagnetized and collisionless. In the limit $k_\perp V_{thi}\sim\mid\omega\mid$, ETG mode resonates with background ions, which deviates from Boltzmannian.
\begin{equation} 
\tilde{n}=-\tau_i \tilde{\phi} [\ 1+i \pi^{\frac{1}{2}} \frac{\omega}{k_y V_{thi}} 
\exp(\ -\frac{\omega^2}{k_y^2 V_{thi}^2} )\ ]\
\label{eqn:1}
\end{equation}
This non-adiabatic ion-response induces a particle flux due to phase lag between density and potential fluctuation. 
The particle flux is given by 
\begin{equation}
\Gamma_n= <\ \tilde{v_r} \tilde{n} >\ = \Sigma_k \tilde{v}_{rk} \tilde{n}_k ^{\ast} = -\Sigma_k \frac{k_y}{B}\mid
\tilde{\phi}_k \mid \mid \tilde{n}_{k}\mid \sin\theta_{n \phi}
\label{eqn:2}
\end{equation}
where $\theta_{n \phi}=\theta_n-\theta_{\phi}$ is the cross angle between density and potential fluctuation.
The particle flux expression for ion response is given by eqn \ref{eqn:1} becomes
\begin{equation}
\Gamma_n=\Sigma_k \pi^{1/2} \tau_i n c_e k_y \rho_e (\ \frac{\omega_r}{k_\perp^2 V_{thi}^2} )\
[\ \exp (\ - \frac{\omega_r^2}{k_\perp^2 V_{thi}^2} )\ ]\ \mid \tilde{\phi}_k \mid ^2
\label{eqn:3}
\end{equation}
where $c_e =\sqrt[2]{\frac{T_e}{M_i}}$, $\rho_e$ is electron larmor radius, $\omega_r \approx -
\left[\left(\frac{1}{L_T}- \frac{2}{3}\frac{1}{L_n}\right)c_{e}k_{y}\rho_{e}k_{z}^{2}c_{e}^{2}/\tau_{e}\right]^{1/3}$ 
is real frequency. Clearly the particle flux is negative
because of real frequency is negative for positive $k_y$. The flux is in gnereral proportional to 
$\left(\frac{1}{L_T}- \frac{2}{3}\frac{1}{L_n}\right)^{1/3} $ and survives even when the density profile becomes flat.
In the flat density region flux is proportional to $(1/L_T)^{1/3}|\phi_{k}|^{2}\propto (1/L_T)^{4/3}$ and hence is of purely thermodiffusive nature. A thrmodiffusivity 
can be defined as follows 
\begin{equation}
\begin{aligned}
\chi_T = -\Sigma_k \pi^{1/2} \tau_i n c_e k_y \rho_e T_e(\ \frac{\left[L_T^{2} c_{e}k_{y}
\rho_{e}k_{z}^{2}c_{e}^{2}/\tau_{e}\right]^{1/3}}{k_\perp^2 V_{thi}^2} ) \\
 [\ \exp (\ - \frac{\omega_r^2}{k_\perp^2 V_{thi}^2} )\ ]\ \mid \tilde{\phi}_k \mid ^2
\end{aligned}
\label{eqn:4}
\end{equation} \\ 
Clearly thermodiffusivity is a nonlinear function of $L_T$ and is proportional to 
$(1/L_T)^{2/3}|\phi_{k}|^{2}\propto (1/L_T)^{5/3}$. 
The experimental results are compared with 
theoretical values for cross angle $\theta_{n\phi} $  obtained from Eq~\ref{eqn:1} and for 
flux $\Gamma $ from Eq.~\ref{eqn:2}. For each comparison the value for $\omega$   and for $k_y $  are 
choosen corresponding to peak power of density perturbation in $\omega - k_y $  space. 
As shown in fig~ \ref{fig:flux_phase} , the radial profiles of cross phase and flux thus obtained follows similar trend as
the experimental profiles and has good agreement with each other.  The small quantitative difference between 
theory and experiment could be due to the fact that for theoretical estimate we only consider the mode with 
maximum power while in experimental observation all the modes are contributing to produce the net particle flux 
larger than the theoretical estimates. However the same argument does not seem to hold around the maximum flux 
region where the theoretical estimates corresponding to maximum power mode is larger than experimental observation. 
The electromagnetic particle flux on the other hand is given by
\begin{eqnarray}
 \Gamma_{em} = -\frac{1}{eB}\left\langle \delta J_{\parallel} \delta B_r \right\rangle \propto
 \sum_k i k_y k_{\perp}^2 |A_{\parallel}|^2
 \label{eqn:5}
\end{eqnarray}
which vanishes since the summand is odd on $k_y$ and also the fact that it has no real part. \\
By the use of the transport equation, for electron, written as 
\begin{equation}
\partial n_e / \partial t + \nabla . \Gamma_e = S_e
\label{eqn:6}
\end{equation}
\begin{equation}
(\partial  / \partial t ) (\frac{3}{2} n_e T_e) + \nabla .(\frac{3}{2}T_e \Gamma_e+q_e) = P_e-\dot{W}
\label{eqn:7}
\end{equation}
We can obtained the change in entropy as derived by Horton et. al. \cite{Horton_entropy} due to the anomalous processes is given by 
\begin{equation}
\frac{dS}{dt}=- \int d \bf{ x } \big [\ \Gamma_e. \frac{\nabla n_e}{n_e} +q_e.\frac{\nabla T_e}{T^2_e}+\frac{\dot{W}_e}{T_e} \big ]\
\label{eqn:8}
\end{equation}
Where $S =\int d {\bf {x}} n_e \ln(\frac{T_e^{3/2}}{n_e})$, is the entropy of the system,  $\Gamma_e$ is particle flux, $ q_e $ is thermal flux, and $ \dot{W} $ is rate of change of wave/fluctuation energy density due to resonant interactions with particles. \\
 From the equation \ref{eqn:8} we find that the inward anomalous particle flux $ \Gamma $ leads the reduction in entropy of system which should be compensated by the radial thermal flux $q_e$ is such that the entropy production from anomalous transport should be positive definite.

\section{Summary and Conclusion}
We studied the particle transport due to turbulent fluctuations in the LVPD. The sources of the underlying turbulence has been established to be due to electron temperature gradient driven in the core of the target region of the device. This is ensured by making the target region free from the energetic electrons by using transverse magnetic field with the help of an EEF. Phase velocity, density -potential correlation and turbulence power spectra confirms that the observed turbulence is driven by ETG. Radial profile of turbulent particle flux and density-potential cross phase ($\theta_{n_{e},\phi}$) has been measured. It is found that the net electrostatic flux is negative i.e., radially inward and is order $\Gamma_{es} \sim - 10^{18} m^{-2}-s^{-1}$. The particle flux maximizes in the region where the $\eta_e$ maximum clearly indicative that the flux is due to the fluctuation resulting from ETG. A net particle flux results from the phase difference between the density and potential fluctuation other than $180^{\circ}$. The Radial profile of density-potential cross phase shows that the cross phase angle deviates from $180^{\circ}$ the most where $\eta_e$ is maximum. Turbulence intensity also maximizes roughly at the location where $\eta_e$ is maximum which act in synergy with the cross phase angle to maximize the flux. The experimental cross phase angle and flux has been compared with the cross phase and flux resulting due to the non-adiabatic ion response due to the resonant interaction of the ions with the ETG mode $k_{\perp} V \sim \omega$. The experimental and theoretical results quantitatively follows the same trend across the radius and matches within $20 \%$ with each other.\\
Electromagnetic component of the particle flux has been also measured and has been found to be insignificant compared to level of electrostatic particle flux $\Gamma_{es} \approx 10^{-7} \times \Gamma_{es}$. Hence complete radial profile of the electromagnetic flux is not attempted. The temporal correlation of parallel electron current and radial magnetic fluctuations are found too weak to produce a significant flux. This is consistent with the fact that theoretically the electromagnetic particle flux in ETG turbulence is expected to be zero.  

\section*{References}

\end{document}